\documentclass[prd,eqsecnum,showpacs,nofootinbib,preprintnumbers]{revtex4}

\usepackage{amsmath,amssymb,amsfonts}
\usepackage[dvips]{graphicx,color}

\newcommand{\ma}[1]{{\mathrm{#1}}}
\newcommand{\pa}{{\partial}}
\newcommand{\calN}{{\cal N}}
\newcommand{\calL}{{\cal L}}
\newcommand{\calO}{{\cal O}}
\newcommand{\gzeta}{{}_G \zeta}
\newcommand{\fzeta}{{}_F \zeta}

\begin{document}
\preprint{YITP-12-4}

\title{A general proof of the equivalence between the $\delta N$ and
 covariant formalisms}

\author{Atsushi Naruko}

\affiliation{Yukawa Institute for Theoretical Physics, Kyoto University,
 Kyoto 606-8502, Japan
}

\date{\today}

\begin{abstract}
Recently, the equivalence between the $\delta N$ and covariant formalisms 
 has been shown (Suyama et al. 2012),
 but they essentially assumed Einstein gravity in their proof.
They showed that the evolution equation of the curvature covector
 in the covariant formalism on uniform energy density slicings
 coincides with that of the curvature perturbation in the $\delta N$ formalism
 assuming the coincidence of uniform energy and uniform expansion (Hubble)
 slicings, which is the case on superhorizon scales in Einstein gravity. 
In this short note, we explicitly show the equivalence between
 the $\delta N$ and covariant formalisms without specifying
 the slicing condition and the associated slicing coincidence,
 in other words, regardless of the gravity theory.
\end{abstract}


\maketitle

\section{Introduction}
Thanks to the current high precision measurements of
 the Cosmic Microwave Background (CMB),
 the nature of its temperature anisotropies has been determined to have
 power spectrum with tiny amplitude $(\sim 10^{-5})$,
 nearly scale invariance and almost Gaussian statistics \cite{Komatsu:2010fb}.
These tiny temperature anisotropies are naturally 
 explained by inflation in the very early universe.
Fortunately or unfortunately, there are hundreds or thousands of
 inflation models consistent with the current observations. 
As such, possible deviations from Gaussianity of the statistics of the CMB,
 non-Gaussianities, have recently been attracting the attention of
 cosmologists as a possible means to constrain inflation models,
 should they be detected \cite{Komatsu:2001rj}.
 (See articles in the focus section in CQG \cite{0264-9381-27-12-120301} and
 references therein for recent developments.)
The PLANCK satellite will indeed provide us with more precise constraints on
 the statistics of the temperature anisotropies in the next couple of years. 
To theoretically compute the non-Gaussianity, we need to 
 include the non-linear dynamics of cosmological perturbations.

The simple and straightforward way to handle such non-linear dynamics is
 the standard perturbative approach
 \cite{Acquaviva:2002ud,Malik:2003mv,Noh:2004bc,Lyth:2005du},
 which can in principle deal with the most general of situations,
 as long as the perturbation expansion is applicable. 
However, the equations can become very much involved and
 quite often the physical transparency may be lost.
To resolve this problem, we have two alternative approaches:
 the $\delta N$ formalism
 \cite{Starobinsky:1986fxa,Sasaki:1995aw,Sasaki:1998ug,Lyth:2004gb}
 and the covariant formalism
 \cite{Langlois:2005ii,Langlois:2005qp,Langlois:2006vv,Langlois:2008vk,
RenauxPetel:2008gi} (see also the review \cite{Langlois:2010vx}).

The $\delta N$ formalism corresponds to the leading order approximation of
 the spatial gradient expansion approach
 \cite{Salopek:1990jq,Deruelle:1994iz,Nambu:1994hu,Shibata:1999zs,
 Lyth:2005du,Yokoyama:2007uu,Yokoyama:2007dw,
 Tanaka:2006zp,Tanaka:2007gh,Takamizu:2008ra,Takamizu:2010xy}.
In the gradient expansion approach,
 the field equations are expanded in powers of spatial gradients and
 hence it is applicable only to perturbations on very large spatial scales.
However, a big advantage is that at leading order in the gradient expansion,
 which corresponds to the separate universe approach \cite{Wands:2000dp},
 the field equations become ordinary differential equations
 with respect to time; hence, the physical quantities at each spatial point
 (where ``each point'' corresponds to a Hubble horizon size region)
 evolve in time independently from those in the rest of the space.
By solving the fiducial homogeneous equation, we can evaluate
 the non-linear dynamics of perturbations and evaluate the generated
 non-Gaussianity. 

One of the most important results obtained in the gradient expansion approach
 is that the full nonlinear curvature perturbation on uniform energy density
 slices is conserved at leading order in the gradient expansion
 if the pressure is only a function of the energy density \cite{Lyth:2004gb}
 (i.e. the perturbation is purely adiabatic). 
This is shown only by using the energy conservation law
 without specifying the gravity theory
 as long as the energy conservation law holds.
Thus, without solving the gravitational field equations,
 one can predict the spectrum and the statistics of the curvature perturbation
 at horizon re-entry during the late radiation or matter-dominated era
 once one knows these properties of the curvature perturbation
 at horizon exit during inflation.
(For Galileon or kinetic braiding models
 \cite{Nicolis:2008in,Deffayet:2010qz,Kobayashi:2010cm, Mizuno:2010ag,
DeFelice:2011zh,Kobayashi:2011pc,DeFelice:2011jm,Kobayashi:2011nu,
Germani:2011ua,Gao:2011qe,DeFelice:2011uc,RenauxPetel:2011uk,Tsujikawa:2012mk},
 this does not hold \cite{Naruko:2011zk,Gao:2011mz})

In the covariant formalism, all quantities are defined
 in a covariant manner and their evolution equations
 are also obtained in a fully non-linear and covariant form.
Since these quantities behave as tensors, they are much easier to
 intuitively understand from a geometrical point of view.
It is known that the curvature covector is conserved
 provided the pressure is purely adiabatic.
One of the main differences from the gradient expansion approach is that
 the covariant approach can be applied to the non-linear dynamics
 at all scales.

However, the relation between quantities in the covariant formalism and 
 those in perturbation theory or the $\delta N$ formalism is unclear,
 since all quantities in the covariant formalism are defined in a covariant way.
For example, although the correspondence between the curvature covector
 and the curvature perturbation is revealed perturbatively
 \cite{Langlois:2005ii,Langlois:2005qp,Enqvist:2006fs,Lehners:2009ja},
 it is not well-understood at full non-linear level
 (see also \cite{Langlois:2008vk}).
The bispectrum of the curvature perturbations in two-field
 inflation models has been studied in terms or these two different formalism
 and their coincidence has been shown in \cite{Watanabe:2011sm}.
More recently, it has been shown that the evolution equation of
 the curvature covector on uniform energy density slicings coincides with
 that of the curvature perturbation in the $\delta N$ formalism,
 assuming the coincidence of uniform energy and uniform expansion (Hubble)
 slicings \cite{Suyama:2012wi}.
Although such slicing coincidence happens at least in Einstein gravity,
 we do not know whether it holds or not in other gravity theories. 
For example, uniform energy and uniform expansion slicings do not coincide
 in $f(R)$ gravity.

In this short note, we give a general proof of the equivalence between
 the $\delta N$ and covariant formalisms without specifying
 the slicing condition and the associated slicing coincidence,
 in other words, independent of the gravity theory.
Here we notice that the $\delta N$ formalism in linear theory,
 which was first proposed in \cite{Sasaki:1995aw},
 gives the final curvature perturbation
 on the uniform energy density slice.
In that paper, they assumed the final time to be some time after 
 complete reheating when the curvature perturbation becomes constant,
 which means the curvature perturbation in the $\delta N$ formalism
 does not depend on the choice of final time, i.e. it is just a constant.
As such, in the case of the original $\delta N$ formalism,
 there is no evolution equation for the curvature perturbation.
However, we do have an evolution equation for
 the curvature perturbation in the lowest-order gradient expansion.
In light of these facts, first, we show that
 the spatial component of the curvature covector
 on uniform energy density slicings in the covariant formalism
 coincides with the spatial gradient of
 the curvature perturbation on the same slicing
 as given by the $\delta N$ formalism.
Next, we show that the evolution equation of the curvature covector 
 is equivalent to the evolution equation of the curvature perturbation
 at the lowest order in the gradient expansion without specifying
 the slicing condition and slicing coincidence or the gravity theory.

\section{Brief reviews}
\label{s:review}

\subsection{Gradient expansion approach and $\delta N$ formalism}
In this subsection, we focus on the dynamics on superhorizon scales
 and use the spatial gradient expansion approach,
 which will be valid on large scales.
We attach a fiducial parameter $\epsilon$ to each spatial derivative
 and expand equations in powers of $\epsilon$.
The lowest order in the gradient expansion corresponds to
 the $\delta N$ formalism.

We express the metric in the ADM form
\begin{gather}
 ds^2 = - \alpha^2 dt + \hat{\gamma}_{i j} (dx^i + \beta^i dt)
 (dx^j + \beta^j dt) \,,
\end{gather}
 where $\alpha$, $\beta^i$ and $\hat{\gamma}_{i j}$ are the lapse function,
 the shift vector and the spatial metric, respectively.
We further decompose the spatial metric as
\begin{gather}
 \hat{\gamma}_{i j} (t, x^k)
 = a^2 (t) e^{2 \psi (t, x^k)} \gamma_{i j} (t, x^k) \,,
 \qquad \det \gamma_{i j} = 1 \,,
\end{gather}
 where $a (t) e^{\psi (t,x^k)}$ is the scale factor at each local point,
 while $a (t)$ is the scale factor of a fiducial homogeneous universe.
By virtue of the separate universe assumption \cite{Wands:2000dp},
 the shift vector is order of $\epsilon$ or higher and
 $\pa_t \gamma_{i j} = \calO (\epsilon^2)$ \cite{Sasaki:1995aw,Lyth:2004gb}.
Then at leading order in gradient expansion,
 we identify $\psi$ as the nonlinear curvature perturbation \cite{Lyth:2004gb}.

Under the separate universe assumption,
 the energy-momentum tensor will take the perfect fluid form
 \begin{gather}
 T_{\mu \nu} = (\rho + P) u_\mu u_\nu + P g_{\mu \nu} + \calO (\epsilon^2) \,,
 \end{gather}
 where $\rho = \rho (t, x^i)$ and $P = P (t, x^i)$ are the energy density
 and pressure of the fluid.
Here we choose the comoving threading as the choice of spatial coordinates
\begin{gather}
 v^i = \frac{u^i}{u^0} \left( = \frac{d x^i}{d t} \right) = 0 \,.
\label{comove}
\end{gather}
 which means the spatial coordinates are chosen so as to
 comove with the fluid.
Then the time component of $u^\mu$ is determined
 by the normalisation condition of $u^\mu$, $u^\mu u_\mu = - 1$, as
\begin{gather}
 u^0 = \frac{1}{\alpha} + \calO (\epsilon^2) \,,
\end{gather}
 where we have used the condition $\beta^i = \calO (\epsilon)$.
The expansion of $u^\mu$ is given by
\begin{gather}
 \Theta \equiv \nabla_\mu u^\mu
 = \frac{3}{\alpha} \left( \frac{\pa_t a}{a} + \pa_t \psi \right)
 + \calO (\epsilon^2) \,.
\label{expansion}
\end{gather}

Next, we write down the local energy conservation equation 
\begin{gather}
 - u_\mu \nabla_\nu T^{\mu \nu}
 = u^\mu \nabla_\mu \rho + \Theta (\rho + P) = 0 \,.
\label{eq-ene-full}
\end{gather}
By inserting Eq (\ref{expansion}), this gives
\begin{gather}
 \frac{\pa_t \, a}{a} + \pa_t \psi
 = - \frac{\pa_t \rho}{3 (\rho + P)} + \calO (\epsilon^2) \,.
\label{eq-evo-Psi}
\end{gather}

In passing, let us choose the uniform energy density slicing. 
From Eq (\ref{eq-evo-Psi}), if $P$ is a function of $\rho$, $P = P (\rho)$,
 the RHS is independent of spatial coordinates because
 it is given by a function of $\rho$, which is a function of time
 on uniform energy density slices.
And hence the curvature perturbation is conserved,
\begin{gather}
 \pa_t \psi_E (t, x^i) = \calO (\epsilon^2) \,, ~~~~~
 \ma{if} ~ P = P (\rho) \,,
\end{gather}
where the subscript $E$ denotes a quantity evaluated
 on the uniform energy density slice.

Finally, we define the $e$-folding number along an integral curve of $u^\mu$:
\begin{gather}
 \calN (t_F, t_I ; x^i)
 \equiv \frac{1}{3} \int_{t_I}^{t_F} d \tau \, \Theta (t, x^i) \,,
\label{calN}
\end{gather}
 where the integral is performed along a constant spatial coordinate line
 because of the condition of comoving threading Eq (\ref{comove}).
By rewriting $\Theta$ with Eq (\ref{expansion}), we have
\begin{align}
 \calN (t_F, t_I, x^i)
 &= \int_{t_I}^{t_F} d t (H + \pa_t \psi) \notag\\
 &= N (t_F, t_I) + \Bigl[ \psi (t_F, x^i) - \psi (t_I, x^i) \Bigr] \,,
\label{calN-Psi}
\end{align}
 where we have introduced the $e$-folding number of
 a fiducial homogeneous universe
 \begin{gather}
 N (t_F, t_I) \equiv \int_{t_I}^{t_F} dt \frac{\pa_t \, a (t)}{a (t)}
 = \log \left[ \frac{a (t_F)}{a (t_I)} \right] \,.
 \end{gather}
From Eq (\ref{calN-Psi}), we see that the difference between
 the actual $e$-folding number $\calN$ and the background value $N$
 gives the change in $\psi$. 
Let us consider two slicings, slicings A and B, which coincide
 at $t = t_I$.
The slicing A is such that it starts on a flat slicing at initial time
 $t_I$ and ends on a uniform energy density slicing at final time $t_F$.
On the other hand, the slicing B is the flat slicing all the time from 
 $t = t_I$ to $t = t_F$.
Then, we have
\begin{align}
 \psi_E (t_F, x^i) &= \calN_A (t_F, t_I ; x^i) - \calN_B (t_F, t_I ; x^i)
 \notag\\
 &= \calN_A (t_F, t_I ; x^i) - N (t_F, t_I)
 \equiv \delta N (t_F, t_I ; x^i) \,,
\label{formula}
\end{align}
 where we have used a property of the flat slicing that
 the $e$-folding number coincides with that of the homogeneous universe
 on the flat slicing.
This is the non-linear $\delta N$ formula.

\subsection{Covariant formalism}
As we have seen in the last subsection,
 the $\delta N$ formalism is a coordinate-based approach
 focusing on the superhorizon dynamics.
On the other hand, in the covariant formalism
 \cite{Langlois:2005ii,Langlois:2005qp},
 all quantities are covariantly defined and
 this formalism is applicable to all length scales.

First, we define the curvature covector, which is one of the most important
 quantity in the covariant formalism,
\begin{gather}
 \zeta_\mu \equiv \pa_\mu \calN - \frac{\dot{\calN}}{\dot{\rho}} \pa_\mu \rho
 \,.
 \label{cc}
\end{gather}
 where the dot on a scalar quantity denotes the derivative along $u^\mu$,
 $\dot{\rho} \equiv u^\mu \nabla_\mu \rho$.
It is also useful to see the relation between $\dot{\calN}$ and $\Theta$
 from the definition of $\calN$, Eq (\ref{calN}),
\begin{gather}
 \dot{\calN} \equiv u^\mu \nabla_\mu \calN = \frac{1}{3} \Theta \,.
\label{deri-calN}
\end{gather}

Here we take the partial derivative of Eq (\ref{eq-ene-full})
 and the equation is rewritten in terms of vector quantities,
 $\rho_\mu (\equiv \pa_\mu \rho)$ and $\calN_\mu (\equiv \pa_\mu \calN)$,
\begin{gather}
 \dot{\rho}_\mu + 3 \dot{\calN} \pa_\mu (\rho + P)
 + 3 \dot{\calN}_\mu (\rho + P) = 0 \,, 
\label{eq-enepa}
\end{gather}
 where the dot on a vector quantity denotes the Lie derivative of its vector
 with respect to $u^\mu$,
 \begin{gather}
 \dot{V}_\mu \equiv \calL_u V_\mu
 \equiv u^\nu \pa_\nu V_\mu + V_\nu \pa_\mu u^\nu \,.
 \end{gather}
After some calculation, we can rewrite Eq (\ref{eq-enepa}) as,
\begin{align}
 \dot{\zeta}_\mu
 &= - \frac{\Theta}{3 (\rho + P)} \left( \pa_\mu P
 - \frac{\dot{P}}{\dot{\rho}} \pa_\mu \rho \right) \,.
\label{eq-evo-cc}
\end{align}
If the pressure is given by a function of the energy density, $P = P (\rho)$,
 the right hand side apparently vanishes and the curvature covector is
 conserved. 

Before closing this subsection, we leave a brief comment on
 the initial condition of the curvature covector.
In the covariant formalism, the $e$-folding number
 is defined as the integration of the expansion
 along an integral curve of $u^\mu$ as in Eq (\ref{calN}).
Then, the derivative of $\calN$ with respect to $u^\mu$,
 Eq (\ref{deri-calN}), is well-defined and gives the expansion.
On the other hand, how can we evaluate the partial derivative of
 the $e$-folding number, $\pa_\mu \calN$, in Eq (\ref{cc}) ?
Since we do not specify the initial time of the integration,
 the $e$-folding numbers for each integral curve cannot be directly compared.
This means the curvature covector has an arbitrariness
 in the choice of the initial hypersurface,
 which we thus need to specify before being able to discuss the relation
 with the curvature perturbation in the $\delta N$ formalism, for example.
Hereinafter, we introduce a prefix for $\zeta_\mu$ to make clear
 the dependence of the curvature covector on the initial slicing.
In particular, ${}_G \zeta_\mu$ and ${}_F \zeta_\mu $ are defined as
 the curvature covector on the general initial slice and
 the initial flat slice respectively,
 \begin{align}
 \gzeta_\mu &\equiv ~ \zeta_\mu ~\ma{on ~ the ~ general ~ initial ~ slice} \,,
 \\ 
 \fzeta_\mu &\equiv ~ \zeta_\mu ~\ma{on ~ the ~ initial ~ flat ~ slice} \,.
 \end{align}

\section{Equivalences}
\label{s:equiv}

\subsection{$\delta N$ formula and the curvature covector}
In this subsection, the relation between the curvature covector
 and the curvature perturbation in the $\delta N$ formalism is revealed.
In particular, we show that the spatial component of
 the curvature covector exactly coincides with the spatial derivative of
 the curvature perturbation on the uniform energy density slicing
 given by the $\delta N$ formalism
 when the initial conditions for the curvature covector are given
 on an appropriate slicing, i.e. flat slicing.
As we have seen in the last subsection,
 we need to specify the initial hypersurface for $\zeta_\mu$
 to compare with the curvature perturbation.
Of course, since the initial hypersurface for $\zeta_\mu$ can be chosen
 arbitrarily, the spatial component of the curvature covector
 does not, in general, coincide with the spatial derivative of
 the curvature perturbation in the $\delta N$ formalism.
This is because the initial hypersurface in the $\delta N$ formalism
 is chosen to be a flat one.

Now we show the equivalence between the $\delta N$ formula and
 the curvature covector. 
From Eq (\ref{cc}), the spatial component of the curvature covector
 on the uniform energy density  slicing is given by the spatial derivative of
 the $e$-folding number, which corresponds to the difference
 in the curvature perturbation on initial and final slices
 \begin{gather}
 \gzeta_i \bigr|_E (t, x^j) = \pa_i \calN_E (t, x^j)
 = \pa_i \Bigl[ \psi_E (t, x^j) - \psi(t_I, x^j) \Bigr] \,, 
\label{rel-cc}
 \end{gather}
 where $\psi_E$ means the curvature perturbation
 on uniform energy density slices.
By appropriately choosing the initial slicing to be a flat slicing, 
 we can further rewrite the above equation as 
 \begin{gather}
 \fzeta_i \bigr|_E (t, x^j) 
 = \pa_i \psi_E (t, x^j) 
 = \pa_i \Bigl[ \delta N (t, t_I ; x^j) \Bigr] \,. 
 \end{gather}
This apparently shows the equivalence between
 the $\delta N$ formula Eq (\ref{formula}) and the curvature covector.
However, if the initial hypersurface is not chosen to be a flat one,
 the equivalence does not hold.

\subsection{Lowest order gradient expansion and the covariant formalism}
Next, we show the equivalence between the lowest order gradient expansion and
 the covariant formalism without specifying the time slicing at final time,
 that is, in a gauge invariant way (see also \cite{Suyama:2012wi}).

First, the time component of curvature covector is expressed as
\begin{gather}
 \gzeta_0 = \pa_t \calN + \frac{\pa_t \rho}{3 (\rho + P)}
 = \calO (\epsilon^2) \,,
\label{zeta0}
\end{gather}
 and the spatial component is
\begin{gather}
 \gzeta_i = \pa_i \calN + \frac{\pa_i \rho}{3 (\rho + P)}  
 = \pa_i \left( \int dt \pa_t \psi \right)
 + \frac{\pa_i \rho}{3 (\rho + P)} \,. 
\label{zetai}
\end{gather}
The evolution equation for the spatial component of the curvature covector
 is given by the spatial part of Eq (\ref{eq-evo-cc})
\begin{align}
 {}_G \dot{\zeta}_i 
 &= - \frac{\Theta}{3 (\rho + P)} \left( \pa_i P
 - \frac{\dot{P}}{\dot{\rho}} \pa_i \rho \right) \notag\\
 &= \frac{1}{\alpha} \frac{\pa_t \rho}{3 (\rho + P)^2} \left( \pa_i P
 - \frac{\pa_t P}{\pa_t \rho} \pa_i \rho \right) \,,
\label{eq-evo-zetai}
\end{align}
 where the explicit form of the LHS is given by
\begin{align}
 {}_G \dot{\zeta}_i
 &= u^\mu \pa_\mu (\gzeta_i) + \gzeta_\mu \pa_i u^\mu 
 = u^0 \pa_t (\gzeta_i) + \gzeta_0 \pa_i u^0 \notag\\
 &= \frac{1}{\alpha} \left[ \pa_t (\gzeta_i)
 - \gzeta_0 \frac{\pa_i \alpha}{\alpha} \right] \,.
\end{align}
By inserting Eq (\ref{zetai}), this reduces to
\begin{align}
 {}_G \dot{\zeta}_i 
 &= \frac{1}{\alpha} \pa_t (\gzeta_i)
 + \calO (\epsilon^3) \notag\\
 &= \frac{1}{\alpha} \left[ \pa_t (\pa_i \psi)
 + \frac{\pa_t \pa_i \rho}{3 (\rho + P)}
 - \frac{\pa_i \rho \, \pa_t (\rho + P)}{3 (\rho + P)^2} \right]
 + \calO (\epsilon^3) \,,
\end{align}
 where we have used Eq (\ref{zeta0}) in the first equality.
Finally, Eq (\ref{eq-evo-zetai}) is rewritten as
\begin{align}
 \pa_i (\pa_t \psi)
 &= \frac{\pa_t \rho}{3 (\rho + P)^2} \left( \pa_i P
 - \frac{\pa_t P}{\pa_t \rho} \pa_i \rho \right) \notag\\
 &\qquad - \frac{\pa_t \pa_i \rho}{3 (\rho + P)}
 + \frac{\pa_i \rho \, \pa_t (\rho + P)}{3 (\rho + P)^2}
 \notag\\
 &= - \pa_i \left[ \frac{\pa_t \rho}{3 (\rho + P)} \right] \,.
\label{eq-evo-zetai-long}
\end{align}
By integrating Eq (\ref{eq-evo-zetai-long}) over $x^i$, we have
\begin{gather}
 \pa_t \psi + C (t) = - \frac{\pa_t \rho}{3 (\rho + P)} \,,
\label{eq-evo-Psi-long}
\end{gather}
 where $C (t)$ is a integration constant. 
By appropriately choosing $C (t)$ so that it coincides with $\pa_t a / a$,
 we can see the coincidence of Eq (\ref{eq-evo-Psi-long}) 
 with Eq (\ref{eq-evo-Psi}).
This shows the equivalence between the lowest order gradient expansion
 and the covariant formalism.

\section{Conclusion}
\label{s:conclusion}
In this short note, we have given a general proof of the equivalence between
 the $\delta N$ and covariant formalisms without specifying
 the slicing condition and the associated slicing coincidence, in other words,
 regardless of the gravity theory.
First, we have shown that the spatial component of the curvature covector on
 the uniform energy density slicing coincides with the spatial gradient of 
 the curvature perturbation on the same slice which is given by
 the $\delta N$ formalism when the initial hypersurface for
 the curvature covector is appropriately chosen so as to be a flat slicing.
Next, we have shown that the evolution equation of the curvature covector 
 is equivalent to the evolution equation of the curvature perturbation
 at the lowest order in the gradient expansion without specifying
 a slicing condition and slicing coincidence or the gravity theory.

In passing we have noted that there is an implicit
 initial slice dependence in the covariant formalism.
Therefore we have emphasised that the equivalence can be shown
 only after the initial slice has been clearly specified.

\acknowledgements
The author gratefully acknowledges Misao Sasaki for fruitful discussions,
 his careful reading of the manuscript and useful suggestions
 and David Langlois for valuable discussions.
This work is supported in part by Monbukagaku-sho Grant-in-Aid
 for the Global COE programs, ``The Next Generation of Physics,
 Spun from Universality and Emergence'' at Kyoto University.
The author is also supported by Grant-in-Aid for JSPS Fellows No.~21-1899.

\end{document}